\documentclass[12pt]{article}
\def\be{\begin{equation}}
\def\ee{\end{equation}}
\def\bea{\begin{eqnarray}}
\def\eea{\end{eqnarray}}

   \def%
\haf{{\frac{1}{2}}} \def\tr{{\rm Tr}} \def\'  \def%
\tphi{\tilde{\phi}}   \def\=\beta 
    
   \def\h1{\hspace{1cm}} 
 \def%
\ddbar{\bar{\Delta}_{[2k] \times [N+2k]}} \def\u U \def%
\ubar{\bar{U}_{[N] \times [N+2k]}}  \def%
\goal{\alpha'\rightarrow 0} \def\ads\def\ola{\overline {\lambda}} \def%
\oep{\overline {\epsilon}}

\begin{document}

\begin{titlepage}
\vskip 2.00cm
\title{
\vbox to\headheight{\hbox to\hsize{\hfill\small\em{\bf BSU-IPP-02/06}}
\vspace{1.5mm}
\footnotesize\hbox to\hsize{\hfill\small\em\ {November~}\number\year~%
}}%
\vspace{2cm}
\bf  Diagonalization of the Hamiltonian in a Chromomagnetic Field}
\author{Sh. Mamedov\thanks{Email: shahin@theory.ipm.ac.ir \hspace{1mm} \&
\hspace{1mm}  Sh$_-$Mamedov@bsu.az}\\
{\small {\em Institute for Physical Problems, Baku State University, }}\\
{\small {\em Z.Khalilov st. 23, AZ-1148, Baku, Azerbaijan}}}
\date{}

\maketitle
\begin{abstract}
We demonstrate  how to find both the color diagonal form of the squared Dirac
 equation in the axial color background and the transformation of the color space,
 which makes this equation diagonal. 
\end{abstract}

\vspace{2cm}
PACS: 03.65.-w 

\end{titlepage}

{\Large {\bf Introduction}} \vspace{4mm} \newline

Classical and Quantum Mechanics of non-abelian charged particles are studied
during a long period. In Refs. [1] and [2] the motion of a colored particle
in the constant color fields in the framework of Quantum Mechanics was
considered . The squared Dirac equation for this motion was solved in the
finite space and the spectrum and wave functions for the considered cases
were found. This equation, in addition to non-diagonal spin matrices,
contains the non-diagonal color matrices as well. Obviously, this
non-diagonality does not permit us to write an eigenvalue equation
separately for the each color state. To this end we need in the diagonal
form of the Hamiltonian in the color space. Having the explicit form of the
Hamiltonian and knowing its hermicity, its diagonal form of it and the
transformation of the color space, which diagonalizes Hamiltonian in this
space can be found. We are going to realize this idea for the case of the
axial chromomagnetic background, which was applied for the study of motion
of the colored particle [1] and supersymmetry of the Dirac equation [9, 3,
7].

\section{Dirac equation in an axial field}

\setcounter{equation}{0}

The Dirac equation for a colored particle in an external color field has the
form: 
\begin{equation}
\label{1.1}\left( \gamma ^\mu P_\mu -M\right) \psi =0, 
\end{equation}
where $P_\mu =p_\mu +gA_\mu =p_\mu +gA_\mu ^a\lambda ^a/2$; the $\lambda ^a$
are Gell-Mann matrices and the color index $a$ runs $a=\overline{1,8}$. $g$
is the color interaction constant. In terms of the Majorana spinors $\phi $
and $\chi $ the equation (1.1) has the following form 
\begin{equation}
\label{1.2}\left( \sigma ^iP_i\right) ^2\psi =-\left( \frac{\partial ^2}{%
\partial t^2}+M^2\right) \psi , 
\end{equation}
where the Pauli matrices $\sigma ^i$ describe the particle's spin. In Eq.
(1.2) and afterwards $\psi $ means $\phi $ or $\chi .$ The spinors $\phi $
and $\chi $ have two components corresponding to the two spin states of a
particle $\psi =\left( 
\begin{array}{c}
\psi _{+} \\ 
\psi _{-} 
\end{array}
\right) .$ Each component of $\psi $ transforms under the fundamental
representation of the color group $SU_c(3)$ and has three color components
describing color states of the particle and corresponding to three
eigenvalues of the color spin $\lambda ^3$

\begin{equation}
\label{1.3}\psi _{\pm }=\left( 
\begin{array}{c}
\psi _{\pm }(\lambda ^3=+1) \\ 
\psi _{\pm }(\lambda ^3=-1) \\ 
\psi _{\pm }(\lambda ^3=0)\;\; 
\end{array}
\right) =\left( 
\begin{array}{c}
\psi _{\pm }^{(1)} \\ 
\psi _{\pm }^{(2)} \\ 
\psi _{\pm }^{(3)} 
\end{array}
\right) . 
\end{equation}

We are going to diagonalize Eq. (1.2) for the case of chromomagnetic
background given by the constant vector potentials $A_\mu ^a$ introduced in
[4]. Remind, for giving the axial chromomagnetic field the components of $%
A_\mu ^a$ are choosen as below

\begin{equation}
\label{1.4}A_1^a=\sqrt{\tau }\delta _{1a},\ A_2^a=\sqrt{\tau }\delta _{2a},\
A_3^a=0,\ A_0^a=0, 
\end{equation}
where $\tau $ is a constant and $\delta _{\mu a}$ is the Kroneker symbol.

For this $A_\mu ^a$ the field strength tensor $F_{\mu \nu }^c=gf^{abc}A_\mu
^aA_\nu ^b$ has only one non-zero component

\begin{equation}
\label{1.5}F_{12}^3=g\tau =H_z^3, 
\end{equation}
and Eq. (1.4) gives a constant field directed along the third axes of the
ordinary and color spaces. Here $f^{abc}$ are the structure constants of the 
$SU_c(3)$ group.

In this field Eq. (1.2) splits into the two equations for the $\psi _{\pm }$%
, which have following forms: 
\begin{equation}
\label{1.6}\left[ {\bf p}^2+\frac 12g^2\tau I_2+g\tau ^{1/2}\left(
p_1\lambda ^1+p_2\lambda ^2\mp \frac 12g\tau ^{1/2}\lambda ^3\right) \right]
\psi _{\pm }=\left( E^2-M^2\right) \psi _{\pm }. 
\end{equation}
Here we have set $i\partial \psi /\partial t=E\psi $ and introduced the
color matrix $I_2=\left( 
\begin{array}{ccc}
1 & 0 & 0 \\ 
0 & 1 & 0 \\ 
0 & 0 & 0 
\end{array}
\right) $. Since field (1.5) is directed along the third axis, $\psi _{\pm }$
describe the spin states with the up and down projections of $\sigma ^3$ and
the corresponding Hamiltonians are defined as $H_{\pm }\psi _{\pm }=E^2\psi
_{\pm }$. $H_{\pm }$ have not diagonal color structure because of the
non-diagonal $\lambda ^1$and $\lambda ^2$ matrices and so, the eigenvalue
equation 
$$
H_{\pm }\psi _{\pm }^{(i)}=E^2\psi _{\pm }^{(i)} 
$$
cannot be written for pure color states $\psi _{\pm }^{(i)}$. The explicit
matrix form of the general Hamiltonian in the combined color and spin spaces
is: 
\begin{equation}
\label{1.7}H=\left( 
\begin{array}{cccccc}
{\cal P}^2 & {\cal G}p_{-} & 0 & 0 & 0 & 0 \\ 
{\cal G}p_{+} & {\cal P}^2+{\cal G}^2 & 0 & 0 & 0 & 0 \\ 
0 & 0 & {\cal P}^2 & 0 & 0 & 0 \\ 
0 & 0 & 0 & {\cal P}^2+{\cal G}^2 & {\cal G}p_{-} & 0 \\ 
0 & 0 & 0 & {\cal G}p_{+} & {\cal P}^2 & 0 \\ 
0 & 0 & 0 & 0 & 0 & {\cal P}^2 
\end{array}
\right) . 
\end{equation}
In Eq. (1.7) we have introduced notations ${\cal P}^2${\it $=$}${\bf p}%
^2+M^2 $, ${\cal G}=g\tau ^{1/2},$ $p_{\pm }=p_1\pm ip_2$. In order to deal
with the matrices with less dimensionality, instead of one with dimension $%
6\times 6$ we can consider two Hamiltonians having dimensions $3\times 3$: 
$$
H_{+}=\left( 
\begin{array}{ccc}
{\cal P}^2 & {\cal G}p_{-} & 0 \\ 
{\cal G}p_{+} & {\cal P}^2+{\cal G}^2 & 0 \\ 
0 & 0 & {\cal P}^2 
\end{array}
\right) ,\qquad H_{-}=\left( 
\begin{array}{ccc}
{\cal P}^2+{\cal G}^2 & {\cal G}p_{-} & 0 \\ 
{\cal G}p_{+} & {\cal P}^2 & 0 \\ 
0 & 0 & {\cal P}^2 
\end{array}
\right) . 
$$
These two Hamiltonians correspond to the spin up and spin down states $\psi
_{+}$ and $\psi _{-}$ respectively. The eigenvalue equation, obviously, can
be written only for the diagonal matrix form of the Hamiltonian. For this
purpose we need in a diagonal form of the Hamiltonian $H$ in the combined
spin-color spin space, i.e. in diagonal forms of $H_{\pm }$. Eigenfunctions
of the diagonalized Hamiltonian will describe the states of the colored
particle, which have definite value of energy%
\footnote{In Ref. [3] we call these states the energy states.}. Because the
Hamiltonian is the hermitian matrix, it has diagonal form in the basis of
its eigenfunctions and this diagonal form is unique. We can find diagonal
form of $H^{\prime }$ for the Hamiltonian (1.7) and then its diagonal
elements will correspond to the energy spectrum of the particle. By $%
H^{\prime }$ we will able to write an eigenvalue equation with the
eigenvalues $E_k^2$%
\footnote{Expressions of $\psi _{\pm }^{(i)}$ and $E_{k}$ can be found in Refs. [3, 1, 10].}%
: 
\begin{equation}
\label{1.8}H^{\prime }\Psi ^{\prime }=E_k^2\Psi ^{\prime }. 
\end{equation}
Here $\Psi ^{\prime }$ is the eigenfunction of $H^{\prime }$, which is
different from $\psi _{\pm }^{(1),(2)}.$ In order to get the diagonal $%
H^{\prime }$ from the non-diagonal Hamiltonian $H$ (1.7), we should make
some transformation $U$ in the combined spin-color spin space. The wave
functions $\psi _{\pm }^{(i)}$ will be transformed on this transformation as
well. More precisely, the Hamiltonian (1.7) will get the diagonal form under
some $U$ transformation of the basis vectors of the combined spin-color spin
space. The basis vectors in the combined space are the functions $\psi _{\pm
}^{(i)}$, since these functions are the eigenfunctions of the spin and color
spin operators $\sigma ^i$ and $\lambda ^a$. Under $U$ transformation of the
combined space, the basis vectors transform according to the rule: 
\begin{equation}
\label{1.9}\Psi ^{\prime }=U\Psi ,\qquad \Psi =\left( 
\begin{array}{c}
\psi _{+}^{(1)} \\ 
\psi _{+}^{(2)} \\ 
\psi _{+}^{(3)} \\ 
\psi _{-}^{(1)} \\ 
\psi _{\_}^{(2)} \\ 
\psi _{-}^{(3)} 
\end{array}
\right) . 
\end{equation}
Components of the new basis vector $\Psi ^{\prime }$ will be some
superposition of $\psi _{\pm }^{(i)}$ and will not be the states with the
definite value of projection of color spin. As we have stated above, if $%
H^{\prime }$ is diagonal, then basis vectors of this space are the
eigenvectors of this Hamiltonian. So, these components are the states with
the definite value of the energy. Since $H^{\prime }$ is unique, the
transformation matrix $U$ and the basis vector $\Psi ^{\prime }$ are unique
as well .

\section{Diagonalization of the Hamiltonian}

\setcounter{equation}{0}

Let us consider how much information can be obtained from the hermicity of $%
H $ for the finding $U$ transformation and $H^{\prime }$. Under
transformation (1.9) the Hamiltonian (1.7), as any matrix in this space, is
transformed as: 
\begin{equation}
\label{2.1}U^{-1}HU=H^{\prime }. 
\end{equation}
The difficulty of determining of $H^{\prime }$ is that, two of three
matrices in (2.1) are unknown; we have no explicit form of either $U$ or $%
H^{\prime }$. It turns out possible to find both of these matrices, relying
on their properties. From the hermicity of $H^{\prime }$ we immediately
conclude, that the matrix $U$ should be unitary, in addition to its
uniqueness: 
$$
H^{\prime \dagger }=U^{\dagger }H^{\dagger }U^{-1\dagger }=H^{\prime
}=U^{-1}HU, 
$$
$$
U^{-1\dagger }=U\Rightarrow UU^{\dagger }=1. 
$$
Eq. (2.1) is the basic relation between two Hamiltonian matrices $H$ and $%
H^{\prime }$. Having multiplied from the right hand side by $U$ matrix it
gets form: 
\begin{equation}
\label{2.2}HU=UH^{\prime }, 
\end{equation}
which contains the linear relations between the elements $u_{ij}$ of $U$
matrix. Another important property of the unitary transformation is that, it
does not change determinant and trace of a matrix. In our case this means
that $H$ and $H^{\prime }$, due to unitarity of $U,$ have the same trace and
the same determinant: 
\begin{equation}
\label{2.3}\det H^{\prime }=\det H,\quad \tr H=\tr H^{\prime }. 
\end{equation}
The matrices $H,$ $H^{\prime }$ and $U$ are matrices of dimensionality $%
6\times 6$. It will be easier to apply Eqs. (2.2) and (2.3) for $H_{+}$ and
for $H_{-}$ separately, than for $H$. To this end we have to divide the $U$
matrix into $U_{+}$ and $U_{-}$ parts, which transform $H_{+}$ and $H_{-}$,
then after solving Eq. (2.2) construct the $U$ matrix by means of $U_{+}$
and $U_{-}$ matrices. Let us remark, that since $H_{+}$ and $H_{-}$ differs
from each other and each matrix has unique transformation diagonalizing it,
the $U_{+}$ and $U_{-}$ matrices should be different as well.

According to (2.3) the determinants of $H_{+}$ and $H_{-}$ equal to
determinants of diagonalized ones, i.e. $H_{+}^{\prime }$ and $H_{-}^{\prime
}$. Besides, they have the same values: 
\begin{equation}
\label{2.4}\det H_{\pm }=\det H_{\pm }^{\prime }=\left( {\cal P}^2+{\cal GP}%
_{\bot }+{\cal G}^2/2\right) \left( {\cal P}^2-{\cal GP}_{\bot }+{\cal G}%
^2/2\right) {\cal P}^2=f_1f_2f_3, 
\end{equation}
where $f_i$ denote $f_{1,2}={\cal P}^2\pm {\cal GP}_{\bot }+{\cal G}^2/2,$ $%
f_3={\cal P}^2$ and ${\cal P}_{\bot }=\sqrt{p_{\bot }^2+{\cal G}^2/4},$ $%
p_{\bot }^2=p_1^2+p_2^2.$ Let us construct firstly the $H_{+}^{\prime }$ by
help of (2.4). Determinant of $H_{+}^{\prime }$ is the product of the
diagonal elements: 
\begin{equation}
\label{2.5}\det H_{+}^{\prime }=\det \left( 
\begin{array}{ccc}
h_{11}^{\prime } & 0 & 0 \\ 
0 & h_{22}^{\prime } & 0 \\ 
0 & 0 & h_{33}^{\prime } 
\end{array}
\right) =h_{11}^{\prime }h_{22}^{\prime }h_{33}^{\prime }. 
\end{equation}
We have two products: the product of three factors $f_i$ in (2.4) and the
product of three diagonal elements $h_{ii}^{\prime }$ in (2.5), which are
equal:%
$$
f_1f_2f_3=h_{11}^{\prime }h_{22}^{\prime }h_{33}^{\prime }. 
$$
Also sum of factors $f_i$ in (2.4) is equal to the sum of diagonal elements
of $H_{+}$ and to the sum of $h_{jj}^{\prime }$ in accordance with (2.3):%
$$
\sum\limits_if_i=3{\cal P}^2+{\cal G}^2=\sum\limits_jh_{jj}=\sum%
\limits_jh_{jj}^{\prime }. 
$$
Equalities of the product and sum allows us to assume that factors $f_i$ in
(2.4 ) are the same with the diagonal elements $h_{jj}^{\prime }$ in (2.5),
since we know invariance of $\tr H_{+}$ and $\det H_{+}$ under $U_{+}$
transformation. But, we encounter the question of place of $f_i$ along the
diagonal of $H_{+}^{\prime }$, i.e. which factor $f_i$ corresponds to the
which diagonal element $h_{ii}^{\prime }$, since we have $3!=6$ different
variants for this corresponding. Each variant can be offered as a candidate
for $H_{+}^{\prime },$ while we have only one $H_{+}^{\prime }.$ Due to
uniqueness of the diagonal form of the hermitian matrix , only one of the
constructed variants will give us the correct $H_{+}^{\prime }$. We may
choose some variant for $f_i\equiv h_{jj}^{\prime }$ identification and then
verify this choice by means of (2.2). According to the uniqueness of $%
H_{+}^{\prime }$ and $U$ the only correct variant of identification will
satisfy Eq. (2.2), i.e. will be solved for the $u_{ij}$ without mathematical
nonsense. We intuitively make the identification below: 
\begin{equation}
\label{2.6}h_{11}^{\prime }\equiv f_2={\cal P}^2-{\cal GP}_{\bot }+{\cal G}%
^2/2,\ \ h_{22}^{\prime }\equiv f_1={\cal P}^2+{\cal GP}_{\bot }+{\cal G}%
^2/2,\ \ h_{33}^{\prime }\equiv f_3={\cal P}^2 
\end{equation}
and Eq. (2.2) for this choice has got the following explicit form: 
$$
\left( 
\begin{array}{ccc}
{\cal P}^2 & {\cal G}p_{-} & 0 \\ 
{\cal G}p_{+} & {\cal P}^2+{\cal G}^2 & 0 \\ 
0 & 0 & {\cal P}^2 
\end{array}
\right) \left( 
\begin{array}{ccc}
u_{11} & u_{12} & u_{13} \\ 
u_{21} & u_{22} & u_{23} \\ 
u_{31} & u_{32} & u_{33} 
\end{array}
\right) = 
$$
$$
=\left( 
\begin{array}{ccc}
u_{11} & u_{12} & u_{13} \\ 
u_{21} & u_{22} & u_{23} \\ 
u_{31} & u_{32} & u_{33} 
\end{array}
\right) \left( 
\begin{array}{ccc}
{\cal P}^2-{\cal GP}_{\bot }+{\cal G}^2/2 & 0 & 0 \\ 
0 & {\cal P}^2+{\cal GP}_{\bot }+{\cal G}^2/2 & 0 \\ 
0 & 0 & {\cal P}^2 
\end{array}
\right) . 
$$
From this equality we get the following system of equations: 
\begin{equation}
\label{2.7}\left\{ 
\begin{array}{c}
p_{-}u_{21}=-\left( 
{\cal P}_{\bot }-{\cal G}/2\right) u_{11}\; \\ p_{-}u_{22}=\left( 
{\cal P}_{\bot }+{\cal G}/2\right) u_{12}\;\;\; \\ p_{+}u_{11}=-\left( 
{\cal P}_{\bot }+{\cal G}/2\right) u_{21}\; \\ p_{+}u_{12}=\left( 
{\cal P}_{\bot }-{\cal G}/2\right) u_{22}\;\;\; \\ 
u_{13}=u_{23}=u_{31}=u_{32}=0 
\end{array}
\right. . 
\end{equation}
The first and second equations in (2.7) are equivalent to the third and
fourth ones, respectively. Thus in (2.7) we have four equations for the
eight real unknowns $Re\ u_{ij}$ and $Im\ u_{ij}$, i.e. the system (2.7) is
unsolvable. But it can be supplemented by the relations between $u_{ij}$
following from the unitarity of $U_{+}$ matrix: 
\begin{equation}
\label{2.8}\left\{ 
\begin{array}{c}
u_{11}u_{11}^{*}+u_{12}u_{12}^{*}=1 \\ 
u_{11}u_{21}^{*}+u_{12}u_{22}^{*}=0 \\ 
u_{21}u_{11}^{*}+u_{22}u_{12}^{*}=0 \\ 
u_{21}u_{21}^{*}+u_{22}u_{22}^{*}=1 \\ 
u_{33}u_{33}^{*}=1\;\;\;\;\;\;\;\;\;\;\;\;\;\;\; 
\end{array}
\right. . 
\end{equation}
Besides, the color summetry group of the Hamilonian (1.7) is $SU\left(
3\right) $ and so, $U_{+}$ transformation belongs to this symmetry group.
One more relation following from the unimodularity condition $\det U_{+}=1$
can be added to the equations of (2.8)%
\footnote{We solve (1.19) for the choice $u_{33}=1$.}:%
$$
u_{11}u_{22}-u_{12}u_{21}=1. 
$$
Equations in (2.8) are non-linear relative to these unknowns and solving of
the systems (2.7) and (2.8) together enables us to find only the module of $%
u_{ij}$: 
\begin{equation}
\label{2.9}u_{11}u_{11}^{*}=\frac 12+\frac{{\cal G}}{4{\cal P}_{\bot }},\
u_{12}u_{12}^{*}=\frac 12-\frac{{\cal G}}{4{\cal P}_{\bot }},\
u_{21}u_{21}^{*}=\frac 12-\frac{{\cal G}}{4{\cal P}_{\bot }},\
u_{22}u_{22}^{*}=\frac 12+\frac{{\cal G}}{4{\cal P}_{\bot }}. 
\end{equation}
We can parametrize $u_{ij}$ introducing free angle parameters $\alpha _i$
and $\beta _i$: 
$$
u_{11}=\left( \frac 12+\frac{{\cal G}}{4{\cal P}_{\bot }}\right)
^{1/2}e^{i\alpha _1},\quad u_{22}=\left( \frac 12+\frac{{\cal G}}{4{\cal P}%
_{\bot }}\right) ^{1/2}e^{i\alpha _2}, 
$$
\begin{equation}
\label{2.10}u_{12}=\left( \frac 12-\frac{{\cal G}}{4{\cal P}_{\bot }}\right)
^{1/2}e^{i\beta _1},\quad u_{21}=\left( \frac 12-\frac{{\cal G}}{4{\cal P}%
_{\bot }}\right) ^{1/2}e^{i\beta _2}. 
\end{equation}
It is useful to write the first equation of the system (2.7) in the terms of 
$Re\ u_{ij}$ and $Im\ u_{ij}$: 
$$
\left\{ 
\begin{array}{c}
p_1\cos \beta _2+p_2\sin \beta _2=-p_{\bot }\cos \alpha _1 \\ 
p_1\sin \beta _2-p_2\cos \beta _2=-p_{\bot }\sin \alpha _1\; 
\end{array}
\right. , 
$$
which, in the angle parametrization (2.10), gives simple relation between $%
\alpha _1$ and $\beta _1$: 
\begin{equation}
\label{2.11}\left\{ 
\begin{array}{c}
p_{\bot }\cos \beta _2=-\left( p_1\cos \alpha _1-p_2\sin \alpha _1\right) \\ 
p_{\bot }\sin \beta _2=-\left( p_2\cos \alpha _1+p_1\sin \alpha _1\right) \; 
\end{array}
\right. . 
\end{equation}
Similarly, the second equation of (2.7) relates the parameter $\alpha _2$
with $\beta _1:$ 
\begin{equation}
\label{2.12}\left\{ 
\begin{array}{c}
p_{\bot }\cos \beta _1=p_1\cos \alpha _2+p_2\sin \alpha _2 \\ 
p_{\bot }\sin \beta _1=p_1\sin \alpha _2-p_2\cos \alpha _2\; 
\end{array}
\right. . 
\end{equation}
Having took these relations into account in (2.10) we eliminate $\beta _i$
parameters and thus cut down the number of free parameters up to two. As a
result, $u_{12}$ and $u_{21}$ are expessed in the $\alpha _j$ angles as
well: 
\begin{equation}
\label{2.13}u_{12}=\frac 1{\sqrt{2}}\left( \frac 1{{\cal P}_{\bot }\left( 
{\cal P}_{\bot }+{\cal G}/2\right) }\right) ^{1/2}p_{-}e^{i\alpha _2},\quad
u_{21}=-\frac 1{\sqrt{2}}\left( \frac 1{{\cal P}_{\bot }\left( {\cal P}%
_{\bot }+{\cal G}/2\right) }\right) ^{1/2}p_{+}e^{i\alpha _1}. 
\end{equation}
Unimodularity condition relate $\alpha _1$ and $\alpha _2$ parameteres: $%
\alpha _2=\alpha _1\equiv \alpha ,$ and then the $U_{+}$ matrix depends on
only one free parameter $\alpha $: 
\begin{equation}
\label{2.14}U_{+}=\left( \frac 1{2{\cal P}_{\bot }}\right) ^{1/2}\left( 
\begin{array}{ccc}
\left( {\cal P}_{\bot }+{\cal G}/2\right) ^{1/2}e^{i\alpha } & p_{-}\left( 
{\cal P}_{\bot }+{\cal G}/2\right) ^{-1/2}e^{-i\alpha } & 0 \\ 
-p_{+}\left( {\cal P}_{\bot }+{\cal G}/2\right) ^{-1/2}e^{i\alpha } & \left( 
{\cal P}_{\bot }+{\cal G}/2\right) ^{1/2}e^{-i\alpha } & 0 \\ 
0 & 0 & \left( 2{\cal P}_{\bot }\right) ^{1/2} 
\end{array}
\right) . 
\end{equation}
As is seen from (2.14), the $U_{+}$ matrix does not transform the $\psi
_{+}^{(3)}$. It can be easily checked, that the transformation (2.14)
reduces $H_{+}$ to the $H_{+}^{\prime }$ with the diagonal elements (2.6).
Let us remark, that any other than (2.6) choice of diagonal elements leads
to mathematical nonsense on solving (2.7) and (2.8) jointly. So, we can
state that (2.6) is the unique diagonal form of the $H_{+}^{\prime }$ and
(2.14) is the unique matrix $U_{+}$ diagonalazing $H_{+}$.

Now we can diagonalize the $H_{-}$ Hamiltonian repeating the scheme used for
the $H_{+}$. According to (2.3) and (2.4) we can suppose equality of the
diagonal elements of $H_{-}^{\prime }$ and the factors $f_i$. We make the
following identification of diagonal elements $h_{jj}^{\prime \prime }$ with
the factors $f_i$: 
\begin{equation}
\label{2.15}h_{11}^{\prime \prime }\equiv f_1={\cal P}^2+{\cal GP}_{\bot }+%
{\cal G}^2/2,\ \ h_{22}^{\prime \prime }\equiv f_2={\cal P}^2-{\cal GP}%
_{\bot }+{\cal G}^2/2,\ \ h_{33}^{\prime \prime }\equiv f_3={\cal P}^2. 
\end{equation}
For the choice (2.15) the relation $H_{-}U_{-}=U_{-}H_{-}^{\prime }$ leads
to the following system of equations%
\footnote{We do not denote the elements of $U_{-}$ matrix differently from ones of $U_{+}$, though they differs.}%
: 
\begin{equation}
\label{2.16}\left\{ 
\begin{array}{c}
p_{\_}u_{21}=\left( 
{\cal P}_{\bot }-{\cal G}/2\right) u_{11}\;\;\;\;\; \\ p_{-}u_{22}=-\left( 
{\cal P}_{\bot }+{\cal G}/2\right) u_{12}\;\; \\ p_{+}u_{11}=\left( 
{\cal P}_{\bot }+{\cal G}/2\right) u_{11}\;\;\;\;\; \\ p_{+}u_{12}=-\left( 
{\cal P}_{\bot }-{\cal G}/2\right) u_{22}\;\; \\ 
u_{13}=u_{23}=u_{31}=u_{32}=0\;\; 
\end{array}
\right. . 
\end{equation}
Having solved this system together with (2.8) we find the $U_{-}$ matrix
diagonalizing $H_{-}$:%
$$
U_{-}=\left( \frac 1{2{\cal P}_{\bot }}\right) ^{1/2}\left( 
\begin{array}{ccc}
\left( {\cal P}_{\bot }+{\cal G}/2\right) ^{1/2}e^{i\beta } & -p_{-}\left( 
{\cal P}_{\bot }+{\cal G}/2\right) ^{-1/2}e^{-i\beta } & 0 \\ 
p_{+}\left( {\cal P}_{\bot }+{\cal G}/2\right) ^{-1/2}e^{i\beta } & \left( 
{\cal P}_{\bot }+{\cal G}/2\right) ^{1/2}e^{-i\beta } & 0 \\ 
0 & 0 & \left( 2{\cal P}_{\bot }\right) ^{1/2} 
\end{array}
\right) . 
$$
Accoding to (2.1) the matrix%
$$
{\cal U}=\left( 
\begin{array}{cc}
U_{+} & 0 \\ 
0 & U_{-} 
\end{array}
\right) 
$$
will reduce the Hamiltonian (1.7) to its diagonal form: 
\begin{equation}
\label{2.17}H^{\prime }=\left( 
\begin{array}{cccccc}
h_{11}^{\prime } & 0 & 0 & 0 & 0 & 0 \\ 
0 & h_{22}^{\prime } & 0 & 0 & 0 & 0 \\ 
0 & 0 & h_{33}^{\prime } & 0 & 0 & 0 \\ 
0 & 0 & 0 & h_{11}^{\prime \prime } & 0 & 0 \\ 
0 & 0 & 0 & 0 & h_{22}^{\prime \prime } & 0 \\ 
0 & 0 & 0 & 0 & 0 & h_{33}^{\prime \prime } 
\end{array}
\right) . 
\end{equation}
In spite of ${\cal U}$ matrix is the transformation of the 6-dimensional
spin-color spin space, actually it transforms the color space, because has
quasidiagonal form and does not mix spin indices.

\end{document}